\documentclass[aps,prl,twocolumn,eqsecnum,amsmath,nobibnotes,nofootinbib,preprintnumbers,superscriptaddress]{revtex4-2}
\usepackage[dvipsnames]{xcolor}
\usepackage{graphicx,float}
\usepackage{amssymb,theorem,mathrsfs,times}
\usepackage{bm,multirow,mathtools,dsfont,setspace}
\usepackage{ulem,cancel}
\usepackage[justification=raggedright,singlelinecheck=true]{caption}
\usepackage{subcaption}
\usepackage{ragged2e}
\usepackage{tikz}
\usetikzlibrary{decorations.pathmorphing,calc,arrows.meta,intersections}
\usepackage[colorlinks, linkcolor=blue, anchorcolor=blue, citecolor=blue]{hyperref}

\begin{document}
\title{On the Extended Kerr--Newman--Bertotti--Robinson Spacetime: \\
Two Black Holes and a Naked Singularity in  Bertotti--Robinson Universe}
\preprint{\hfill {\small {ICTS-USTC/PCFT-26-61}}}
\date{\today}

\author{Yu-Sen Zhou}
\email{zhou\_ys@mail.ustc.edu.cn}
\affiliation{Interdisciplinary Center for Theoretical Study and Department of Modern Physics,\\
University of Science and Technology of China, Hefei, Anhui 230026, China}

\author{Wen-Tao Fu}
\email{fuwentao2024@mail.ustc.edu.cn}
\affiliation{Interdisciplinary Center for Theoretical Study and Department of Modern Physics,\\
University of Science and Technology of China, Hefei, Anhui 230026, China}

\author{Li-Ming Cao}
\email{caolm@ustc.edu.cn}
\affiliation{Interdisciplinary Center for Theoretical Study and Department of Modern Physics,\\
University of Science and Technology of China, Hefei, Anhui 230026, China}
\affiliation{Peng Huanwu Center for Fundamental Theory, Hefei, Anhui 230026, China}

\author{Rong-Gen Cai}
\email{caironggen@nbu.edu.cn}
\affiliation{Institute of Fundamental Physics and Quantum Technology, \& School of Physical Science and Technology, Ningbo University, Ningbo 315211, China}

\begin{abstract}
    We introduce new coordinates that extend the Kerr--Newman--Bertotti--Robinson spacetime consistently with the previous reciprocal continuation. In these coordinates, the two points at $\Omega=0$ are resolved into complete Bertotti--Robinson null infinities. The extended spacetime describes two black holes sharing a common exterior of nontrivial topology, together with a naked ring singularity. In the regular static limit, the new coordinates globally identify the spacetime with the balanced Alekseev--Garc\'ia geometry.
\end{abstract}

\maketitle

\section{Introduction}
Astrophysical black holes are generally embedded in electromagnetic environments, and external magnetic fields can substantially affect their geometry and dynamics. Exact black-hole solutions in magnetic backgrounds remain scarce~\cite{Ernst:1976mzr,Ernst:1976bsr,Alekseev:1996fq}, despite providing controlled settings for studying the interplay between gravity, rotation, and electromagnetic fields. An important example is the Kerr--Bertotti--Robinson ($\text{KBR}_\text{s}$) spacetime~\cite{Podolsky:2025tle}, which describes a rotating black hole in a Bertotti--Robinson (BR) background. More recently, this solution has been embedded in the broader Kerr--Newman--Bertotti--Robinson (KNBR) family~\cite{Ovcharenko:2026tos}, incorporating electric charge while retaining the characteristic BR asymptotics. These geometries have attracted considerable attention from several perspectives~\cite{Podolsky:2025tle,Zhou:2026tkm,Ovcharenko:2026ooh,Zhou:2026fvj,Ovcharenko:2025cpm,Wang:2025vsx,Zeng:2025olq,Ali:2025beh,Wang:2025bjf,Zeng:2025tji,Ahmed:2025ril,Siahaan:2025ngu,Li:2025rtf,Mirkhaydarov:2026fyn,Wang:2026czl,Mustafa:2026gly,Siahaan:2026tuf,Lu:2026kcm,Rehman:2026rzq,Hu:2026slp,Wan:2026lca,Ahmed:2026ozq,Roy:2026poj,Hassanabadi:2026fzz,Huang:2026qzj,Singh:2026rbz,Gray:2025lwy,Xu:2026ags,Ovcharenko:2026tos,Kubiznak:2026uro,Wan:2026bkb,Zhang:2026sgj}.

Despite this progress, the location of the null infinity of the KNBR geometry remains obscure in its original Boyer--Lindquist--type coordinates. Null infinity is the asymptotic boundary through which a gravitational system communicates with the external universe, and its structure characterizes the environment in which the system is embedded. In KNBR, however, the apparent boundary $r=\infty$ is not null infinity. Light rays reach it in finite affine parameter, observers arrive there in finite proper time, and the Weyl curvature that characterizes the gravitational field remains nonzero. Thus $r=\infty$ is a regular continuation surface rather than a genuine asymptotic boundary.

Previous analyses exploited this observation by introducing a reciprocal radial coordinate $y\propto 1/r$, which smoothly extends the spacetime from $r=+\infty$ in one Boyer--Lindquist patch to $r=-\infty$ in a neighboring one~\cite{Zhou:2026tkm,Ovcharenko:2026ooh,Zhou:2026fvj}. Repeating this continuation generates an infinite chain of connected regions and identifies the loci $\Omega=0$ as the true null infinity. Although this construction establishes extendibility across $r=\infty$, the global meaning of the continuation, the structure of the resulting null infinity, and possible alternative global interpretations remain unclear.

In this work, we introduce coordinates that reproduce the reciprocal continuation while smoothly joining the two Boyer--Lindquist patches across $r=\infty$ in a single chart. The two points at $\Omega=0$ are thereby resolved into complete BR null infinities. Using these coordinates, we construct a global Weyl orbit diagram~\cite{Harmark:2004rm,Harmark:2005vn} displaying the patch structure, joining surface, horizons, and asymptotic regions. The resulting causal structure reveals that the two Killing horizons, previously identified as the outer horizon of one Boyer--Lindquist patch and the inner horizon of the neighboring patch, in fact form two disconnected components of the future event horizon, describing two black holes that share a common exterior with nontrivial spatial topology. Meanwhile, the ring singularity is causally visible from the BR null infinities and is therefore naked. In the regular static limit, namely the Schwarzschild--BR solution, the coordinates further establish a global identification with the balanced Alekseev--Garc\'ia geometry~\cite{Alekseev:1996fq,Ortaggio:2018ikt}. Their relation had been discussed previously~\cite{Barrientos:2026shy,Ovcharenko:2026byw}, and recently the local equivalence between the two was established by an explicit diffeomorphism~\cite{Astorino:2026nhd}. Details are given in Appendix~\ref{sec:static}.

\section{Extended KNBR spacetime}
The KNBR metric~\cite{Ovcharenko:2026tos}, in Boyer--Lindquist-type coordinates, is
\begin{align}
\mathrm{d}s^2&=\frac{1}{\Omega^2}\Biggl[-\frac{Q}{\rho^2}\left(\mathrm{d}t-\frac{a\sin^2\theta}{P_0}\mathrm{d}\phi\right)^2+\frac{\rho^2}{Q}\mathrm{d}r^2+\frac{\rho^2}{P}\mathrm{d}\theta^2\nonumber\\
&+\frac{P\sin^2\theta}{\rho^2}\left(a\mathrm{d}t-\frac{r^2+a^2}{P_0}\mathrm{d}\phi\right)^2\Biggr]\,,
\label{eq:metric}
\end{align}
where
\begin{align}
\rho^2&=r^2+a^2\cos^2\theta\,, &\Delta&=Dr^2-2mr+(1+k^2)a^2\,,\nonumber\\
I&=(1+kBr)^2+B^2r^2\,, &P&=1+B^2\mu^2\cos^2\theta\,,\nonumber\\
\Omega^2&=I-B^2\Delta\cos^2\theta\,, &Q&=I\Delta\, ,
\label{eq:f}
\end{align}
with
\begin{align}
D&=1+k^2-\frac{e^2}{a^2}\,,\quad k=\frac{es-amB}{a(1+e^2B^2)}\,,\nonumber\\
s^2&=1+(e^2-a^2)B^2-(m^2+e^2)a^2B^4\,,\nonumber\\
\mu^2&=m^2-(1+k^2)a^2D\,.
\label{eq:parameters}
\end{align}
The azimuthal coordinate $\phi$ now has period $2\pi$, since the conicity factor, $P_0\equiv P(0)=1+B^2\mu^2$, is explicitly included in the metric. Here $m$, $a$, $B$, and $e$ denote the mass, rotation, background magnetic field, and black-hole charge parameters, respectively. We first focus on the generic case with $m>0$, $a>0$, $B>0$, $e>0$, $s>0$, $\mu>0$, and $D>0$. The special cases are discussed in Appendix~\ref{sec:special}. The outer and inner Killing horizons are then located at the real positive roots of $\Delta$, $r_1$ and $r_2$, respectively, given by
\begin{equation}
r_{1,2}=\frac{m\pm\mu}{D}\,.
\label{eq:horizons}
\end{equation}
The electromagnetic field is described by the complex potential, which, remarkably, takes the same form as in~\cite{Podolsky:2025tle}:
\begin{align}
\boldsymbol{A}=&-\frac{\mathrm{e}^{\mathrm{i}\nu}}{2B}\Biggl[
\frac{\Omega_{,r}}{\chi}
\left(a\mathrm{d}t-\frac{r^2+a^2}{P_0}\mathrm{d}\phi\right)\nonumber\\
&+\frac{\mathrm{i}\Omega_{,\theta}}{\chi\sin\theta}
\left(\mathrm{d}t-\frac{a\sin^2\theta}{P_0}\mathrm{d}\phi\right)
+\frac{\Omega-1}{P_0}\mathrm{d}\phi
\Biggr]\,,
\label{eq:Ac}
\end{align}
where $\chi\equiv r+\mathrm{i}a\cos\theta$ and $\nu$ is an arbitrary constant duality angle. Here the sign of $\Omega$ may be chosen freely, as changing its sign is equivalent to $\nu\to\nu+\pi$. The associated Maxwell field $\boldsymbol{F}=2\mathrm{d}(\operatorname{Re}\boldsymbol{A})$ vanishes under the joint limit $(e,B)\to(0,0)$, independent of the path. Together with the metric~\eqref{eq:metric}, it solves the Einstein--Maxwell equations.

However, this coordinate system suffers from a significant limitation. The surface $r=\infty$ is not a null infinity. This can be seen from the fact that null geodesics can arrive at $r=\infty$ within a finite affine parameter, timelike geodesic observers can reach it within a finite proper time, and the Weyl scalar $\Psi_2$ does not decay there. Therefore, one can extend the spacetime across this surface by introducing a reciprocal radial coordinate $y\propto1/r$, which connects the $r=+\infty$ side of one patch to the $r=-\infty$ side of a neighboring patch smoothly. Repeating this continuation generates the previously identified infinite chain of connected regions~\cite{Zhou:2026tkm,Ovcharenko:2026ooh,Zhou:2026fvj}. Nevertheless, the true null infinity remains nonmanifest in this extended coordinate description.

We identify coordinates that reveal null infinity as an asymptotically Bertotti–Robinson end end and naturally encode the previously identified extension across $r=\infty$:
\begin{align}
\tau&=\frac{t}{BP_0L^2}\,,&R&=\frac{\cos\theta}{E}\frac{r-r_{\rm c}}{\Omega}\,,\nonumber\\
\cos\Theta&=\frac{Be}{a}\frac{r-r_\Omega}{\Omega}\, ,& \Phi &=\phi-\omega_\Omega t\,,
\label{eq:tr}
\end{align}
where
\begin{align}
r_\Omega&=-\frac{a^2(k+mB)}{Be^2}\,,&r_{\rm c}&=r_\Omega-\frac{2\Delta(r_\Omega)}{\Delta'(r_\Omega)}\,,\nonumber\\
E&=\sqrt{\frac{4\Delta(r_\Omega)}{B^2P_0[\Delta'(r_\Omega)]^2}}\,,&L&=\sqrt{\frac{r_\Omega^2+a^2}{B^2P_0\Delta(r_\Omega)}}\,,
\label{eq:aux}
\end{align}
and
$
\omega_\Omega=\omega(r_\Omega,0)$ with $\omega(r,\theta)\equiv-g_{t\phi}/g_{\phi\phi}$.
Moreover, $\Omega^2$ vanishes only at $r=r_\Omega$ on the axes $\theta=0,\pi$, while $r_2<r_{\rm c}<r_1$. The inverse transformation is given by
\begin{align}
t&=BP_0L^2\tau\,,&r&=r_\Omega+\frac{a(r_{\rm c}-r_\Omega)\cos\Theta}{a\cos\Theta-BeE\xi}\,,\nonumber\\
\cos\theta&=\frac{R}{\xi}\,,&\phi&=\Phi+BP_0L^2\omega_\Omega\tau\, .
\label{eq:inv}
\end{align}
Here
\begin{equation}
\xi\equiv\frac{r-r_{\rm c}}{E\Omega}
\label{eq:zeta}
\end{equation}
can be expressed entirely in terms of $(R,\Theta)$ as $\xi=\sqrt{u}$, where $u$ is the larger root of
$$
u^2-\left(R^2+\sin^2\Theta+B^2\mu^2\right)u+B^2\mu^2R^2\cos^2\Theta=0\,,
$$
chosen to ensure $u\ge R^2$, and hence $|\cos\theta|\le1$.

The $(R,\Theta)$ coordinates actually cover only the $\Delta\geq0$ region, as shown by the identity
\begin{equation}
\sin^2\Theta=\frac{B^2\Delta\sin^2\theta}{\Omega^2}\,.
\label{eq:sth}
\end{equation}
Eq.~\eqref{eq:zeta} then shows that the choice of positive $\xi$ fixes $\operatorname{sgn}\Omega=\operatorname{sgn}(r-r_{\rm c})$ in the $\Delta>0$ region.

In the new coordinates, the metric takes the form
\begin{align}
\mathrm{d}s^2&=\mathcal F\left(\frac{\mathrm{d}R^2}{1+R^2}+\mathrm{d}\Theta^2\right)-\frac{L^4(1+R^2)}{g_{\phi\phi}/\sin^2\Theta}\mathrm{d}\tau^2\nonumber\\
&+g_{\phi\phi}\left[\mathrm{d}\Phi-BP_0L^2(\omega-\omega_\Omega)\mathrm{d}\tau\right]^2\, ,
\label{eq:mRT}
\end{align}
where
$$
\mathcal F\equiv\frac{\rho^2}{B^2\left(P\Delta+\mu^2I\sin^2\theta\right)}\,.
$$
As $R\to\pm\infty$, both $\mathcal F$ and $g_{\phi\phi}/\sin^2\Theta$ approach the common limit $L^2$, while $\omega-\omega_\Omega\to0$, hence the metric (\ref{eq:mRT}) reduces to
$$
 L^2\left[-(1+R^2)\mathrm{d}\tau^2+\frac{\mathrm{d}R^2}{1+R^2}+\mathrm{d}\Theta^2+\sin^2\Theta\,\mathrm{d}\Phi^2\right]\,,
$$
which is the standard BR geometry, $\mathrm{AdS}_2\times S^2$, with radius $L$. The $R\to\pm\infty$ are the two timelike null infinities into which the $\Omega=0$ points of the old coordinates are resolved. Indeed,
\begin{align}
\frac{\sin\theta}{r-r_\Omega}&=-\frac{e}{a\sqrt{\Delta(r_\Omega)}}\tan\Theta+\mathcal{O}(\Omega)\,,\nonumber\\
\Omega&=\frac{r_\Omega-r_{\rm c}}{E}|R|^{-1}+\mathcal{O}(R^{-2})\,.\nonumber
\label{eq:omr}
\end{align}
Thus $\Theta$ resolves the direction of approach to $\Omega=0$, while $|R|$ controls the distance to it.

The hypersurface $r=r_\Omega$ is mapped to the new equatorial plane $\Theta=\pi/2$, where $R=\sqrt{P_0}\cot\theta\,.$
The $r=\infty$ surface becomes the hypersurface $\mathcal{Y}=0$, where
\begin{equation}
\mathcal{Y}\equiv a\cos\Theta-BeE\xi=\frac{a(r_{\rm c}-r_\Omega)\cos\Theta}{r-r_\Omega}\,.
\label{eq:y}
\end{equation}
This hypersurface lies at finite coordinates and is regular in the new coordinate system. Since $\cos\Theta$ approaches a nonvanishing $\theta$-dependent value as $\mathcal{Y}\to0$, one has $\mathcal{Y}\propto1/r$ near this hypersurface, so the previous reciprocal coordinate $y\propto1/r$ is simply an equivalent local transverse coordinate across it.

The horizons $r=r_{1,2}$ become the finite rods $|R|\leq B\mu$ on $\Theta=0,\pi$, respectively. The complementary segments $|R|>B\mu$ on $\Theta=0,\pi$ are the symmetry axes. The pronounced asymmetry between $r_1$ and $r_2$ in the usual description is thus partly a consequence of the Boyer--Lindquist radial organization. In the new coordinates, the two horizons appear on a much more symmetric footing, although they remain intrinsically distinct.

$R=0$ is precisely the old equator $\theta=\pi/2$, with $R>0$ and $R<0$ representing the two hemispheres and connected smoothly across $R=0$. To reveal the geometry of the $R=0$ locus, we introduce the canonical Weyl coordinates~\cite{Harmark:2004rm,Harmark:2005vn} associated with the Killing fields $\partial_\tau$ and $\partial_\Phi$, i.e., 
\begin{equation}
\varrho=L^2\sqrt{1+R^2}\sin\Theta\,,\qquad z=L^2R\cos\Theta\,.
\label{eq:w}
\end{equation}
Together with the auxiliary Cartesian coordinates
\begin{align}
X=\varrho\cos\Phi\,,\quad Y=\varrho\sin\Phi\,,\quad Z=z\, ,
\label{eq:xyz}
\end{align}
the surfaces $R=\mathrm{const}$ satisfy
\begin{equation}
\frac{X^2+Y^2}{1+R^2}+\frac{Z^2}{R^2}=L^4\,.
\label{eq:sph}
\end{equation}
As $R\to0$, these oblate spheroids degenerate to the equatorial disk
\begin{equation}
Z=0\,,\quad X^2+Y^2=L^4\sin^2\Theta\le L^4\,,
\label{eq:disk}
\end{equation}
so $R=0$ is a disk which is compressed from the old equator outside the horizons. The transformation $(R,\Theta)\to(-R,\pi-\Theta)$ leaves $(X,Y,Z)$ unchanged, exhibiting the two-sheeted structure across the disk similar to the $r=0$ disk of Kerr. We denote the two sheets of the disk by $\mathcal{D}_+$ and $\mathcal{D}_-$, approached from $R>0$ and $R<0$, respectively. Thus, in the $(R,\Theta)$ description, an observer entering the disk from the northern hemisphere through $\mathcal{D}_+$ emerges from $\mathcal{D}_-$ into the southern hemisphere, corresponding in the old coordinates to crossing the equatorial plane from north to south. On the disk, one has
\begin{equation}
\cos\Theta=\frac{Be(r-r_\Omega)}{a\,\operatorname{sgn}(r-r_{\rm c})\sqrt{I}}\,.
\label{eq:th0}
\end{equation}
The singularity at $r=0$ and $\theta=\pi/2$ is therefore represented by the ring-shaped locus on $R=0$  satisfying
\begin{equation}
\cos\Theta=\frac{Be r_\Omega}{a}=-\frac{a(k+mB)}{e}\,.
\label{eq:sing}
\end{equation}
The horizon rods intersect the disk at its center, $X=Y=Z=0$. Consider an observer skimming above the old equator, $\theta\to\pi/2^-$, moving toward increasing $r$ from near the outer horizon $r=r_1$. In the $(R,\Theta)$ coordinates, the trajectory starts near the center on the upper face of $\mathcal{D}_+$ and crosses the circle $\mathcal{Y}=0$, where $r=+\infty$ in one $(r,\theta)$ chart is continued as $r=-\infty$ in the neighboring chart. With $r$ continuing to increase, the $(R,\Theta)$ trajectory reaches the rim at $\Theta=\pi/2$, where $r=r_\Omega$, rounds onto the lower face of the same $\mathcal{D}_+$ sheet, and heads back toward the center with $\Theta>\pi/2$. As $r$ passes through zero, the trajectory skirts the ring-shaped singular locus given by Eq.~\eqref{eq:sing}, and eventually reaches the center at $\Theta=\pi$, where the inner-horizon rod terminates.

\begin{figure}[htbp]
    \centering
    \includegraphics[width=0.87\linewidth]{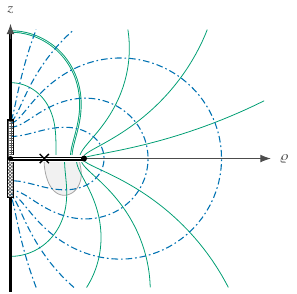}
    \caption{Weyl orbit-space representation of the $R>0$ ($\theta<\pi/2$) sheet. Green solid curves have constant $r$, with the double green curve denoting $r=\infty$, while blue dash-dotted curves have constant $\theta$. The thick portions of the $z$ axis denote symmetry-axis rods, while the upper and lower patterned segments denote the outer and inner horizon rods, respectively. The double line segment on $z=0$ represents the two faces of $\mathcal D_+$ of the $R=0$ disk, while $\mathcal D_-$ resides on the other sheet. The upper face of $\mathcal D_+$ is identified with the lower face of $\mathcal D_-$, and the lower face of $\mathcal D_+$ with the upper face of $\mathcal D_-$. The cross marks the ring singularity on the lower face of $\mathcal D_+$, surrounded by the gray-shaded CTC region where $g_{\phi\phi}<0$ and the axial Killing orbits are timelike, thus the Weyl diagram there is a symmetry-orbit-space representation rather than a spatial cross section. The $R<0$ ($\theta>\pi/2$) sheet is obtained by $z\to-z$, corresponding to $\theta\to\pi-\theta$, with the outer and inner horizon rods exchanged and $\mathcal D_+$ replaced by $\mathcal D_-$.}
    \label{fig:weyl-orbit}
\end{figure}

We now focus on the global structure of the orbit space, which is particularly transparent in the Weyl diagram. In terms of the Weyl coordinates~\eqref{eq:w}, one has
\begin{align}
\mathrm{d}s^2&=\frac{\mathcal F}{L^4(R^2+\cos^2\Theta)}\left(\mathrm{d}\varrho^2+\mathrm{d}z^2\right)-\frac{\varrho^2}{g_{\phi\phi}}\mathrm{d}\tau^2\nonumber\\
&+g_{\phi\phi}\left[\mathrm{d}\Phi-BP_0L^2(\omega-\omega_\Omega)\mathrm{d}\tau\right]^2\,,
\label{eq:mw}
\end{align}
and the Weyl diagram is shown in Fig.~\ref{fig:weyl-orbit}. Each horizon rod has length $2L^2B\mu$, which is equivalently $|\kappa\mathcal A/(2\pi)|=4|TS|$~\cite{Clement:2019ghi}. Remarkably, the two horizon rods have equal lengths. 

\begin{figure}[htbp]
    \centering
    \includegraphics[width=1\linewidth]{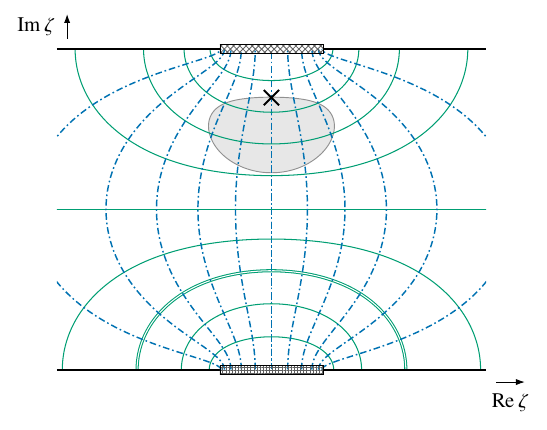}
    \caption{Conformal Weyl representation of the full two-sheeted orbit space in $\zeta=\operatorname{arsinh}R+\mathrm{i}(\Theta-\pi/2)$, which unfolds the branching at the disk rim $\zeta=0$. The two sheets occupy $\operatorname{Re}\zeta>0$ and $\operatorname{Re}\zeta<0$, joined across $\operatorname{Re}\zeta=0$, while $\operatorname{Im}\zeta=\pm\pi/2$ form the axis/horizon boundaries. Curve styles, symbols, and shading are as in Fig.~\ref{fig:weyl-orbit}.}
    \label{fig:conformal-weyl}
\end{figure}

Consider a loop in the $(\varrho,z)$ plane around the disk rim at $(L^2,0)$. Owing to the disk's two-sheeted structure, at least two circuits are required to return to the original spacetime point, reminiscent of the square-root branch structure in the complex plane. Introduce the conformal coordinate
\begin{equation}
\zeta=\operatorname{arsinh}R+\mathrm{i}\left(\Theta-\frac{\pi}{2}\right)\,,\quad z+\mathrm{i}\varrho=\mathrm{i}L^2\cosh\zeta\,,
\end{equation}
which, near the rim at $\zeta=0$, takes the local square-root form
\begin{equation}
\zeta^2=\frac{2}{L^2}\left(\varrho-L^2-\mathrm{i}z\right)+\mathcal O(|\zeta|^4)\,,
\end{equation}
thereby unfolding the two-sheeted branching. The $(R,\Theta)$-space metric becomes
\begin{equation}
\mathrm{d}s_{\rm orb}^2=\mathcal F|\mathrm{d}\zeta|^2\,,
\end{equation}
and the resulting conformal Weyl diagram is shown in Fig.~\ref{fig:conformal-weyl}.

\section{The Global Causal Structure}
\label{sec:causal}

We next show that there are, in fact, two black holes, and a naked singularity. We first establish what is causally visible from null infinity. Denote the Killing horizons at $r=r_1$ and $r=r_2$ by $\mathcal H_1$ and $\mathcal H_2$, respectively. Let $\mathscr E$ denote the common stationary exterior and $\mathscr I=\mathscr I_{+}\cup\mathscr I_{-}$ its two BR asymptotic ends at $R\to\pm\infty$. Every regular point of $\mathscr E$ can send a causal signal to $\mathscr I$. To show this, introduce two useful future-directed null directions
\begin{align}
\ell_\pm&=\frac{r^2+a^2}{Q}\partial_t\pm\partial_r+\frac{aP_0}{Q}\partial_\phi\,,\nonumber\\
n_\pm&=\frac{r^2+a^2}{\sqrt{PQ}}\partial_t\pm\partial_\theta+\frac{aP_0}{\sqrt{PQ}}\partial_\phi\,.
\label{eq:light}
\end{align}
From any $p\in\mathscr E$, one can construct a piecewise-null curve by following an appropriate $\ell_\pm$ toward $r=r_\Omega$, detouring along $n_\pm$ if needed to avoid the singularity, and then following $n_\pm$ asymptotically toward $\Omega=0$. The asymptotic behavior of $P$ and $Q$ near $\mathcal{Y}=0$ and their regularity along each segment ensure finite accumulated shifts in $t$ and $\phi$, after which the piecewise-null trajectory can be smoothed into a causal curve while preserving causality. Thus
\begin{equation}
\mathscr E\subset J^-(\mathscr I)
\label{eq:past}
\end{equation}

The ring singularity is causally visible. On the equatorial plane $\theta=\pi/2$, the integral curves of $I(r)\ell_-$ are affinely parametrized null geodesics. Consider such a geodesic through a regular point with $r<0$. Since $I(0)=1$, the geodesic reaches the ring singularity at finite affine parameter in the past and enters the regular exterior in the future, from which the null infinities are causally accessible. Hence the ring singularity is naked.

It remains to identify the boundary of this causal past. In each stationary region,
\begin{equation}
\mathcal T=\partial_t+\frac{aP_0}{r^2+a^2}\partial_\phi\,,\quad g(\mathcal T,\mathcal T)=-\frac{Q\rho^2}{\Omega^2(r^2+a^2)^2}<0\,,
\end{equation}
defines a time orientation that extends smoothly across the regular horizons and joining surfaces. We choose $\mathcal T$ future directed. The future branches of the two Killing horizons $\mathcal H_1$ and $\mathcal H_2$ are denoted by $\mathcal H_1^+$ and $\mathcal H_2^+$, respectively, with generators
\begin{equation}
K_i=\partial_t+\omega_i\partial_\phi\,,\quad i=1,2\,.
\label{eq:Killing}
\end{equation}
where $\omega_{1,2}\equiv\omega(r_{1,2},\theta)$ is independent of $\theta$.

To determine the causal orientation at $\mathcal H_1^+$, introduce ingoing Eddington--Finkelstein coordinates regular across the future horizon,
\begin{equation}
\mathrm{d}v=\mathrm{d}t+\frac{r^2+a^2}{Q}\mathrm{d}r\,,\quad \mathrm{d}\tilde\phi_1=\mathrm{d}\phi+\frac{aP_0}{Q}\mathrm{d}r\,.
\end{equation}
In these coordinates, one has
\begin{equation}
g_{ab}K_1^b=\frac{\rho_1^2}{\Omega_1^2(r_1^2+a^2)}\partial_a r\quad\text{on }\mathcal H_1^+\,,
\end{equation}
where $\rho_i^2=r_i^2+a^2\cos^2\theta$ and $\Omega_i^2=\Omega^2(r_i,\theta)=I(r_i)$.

Near $\mathcal H_2^+$, the regular future extension instead uses outgoing Eddington--Finkelstein coordinates
\begin{equation}
\mathrm{d}u=\mathrm{d}t-\frac{r^2+a^2}{Q}\mathrm{d}r\,,\quad \mathrm{d}\tilde\phi_2=\mathrm{d}\phi-\frac{aP_0}{Q}\mathrm{d}r\,.
\end{equation}
Therefore, one arrives at
\begin{equation}
g_{ab}K_2^b=-\frac{\rho_2^2}{\Omega_2^2(r_2^2+a^2)}\partial_a r\quad\text{on }\mathcal H_2^+\,.
\end{equation}
Since $K_i$ and any future-directed causal vector $X^a$ satisfy $g(K_i,X)\leq0$, it follows that
\begin{equation}
X^a\partial_a r\leq0\quad\text{on }\mathcal H_1^+,\quad X^a\partial_a r\geq0\quad\text{on }\mathcal H_2^+\,.
\label{eq:one}
\end{equation}
Immediately beyond either horizon,
\begin{equation}
g(\nabla r,\nabla r)=\frac{\Omega^2Q}{\rho^2}<0\,,
\label{eq:time}
\end{equation}
so $r$ is strictly monotone along future-directed causal curves in the corresponding $\Delta<0$ region.

The global time orientation fixes the allowed direction in which future-directed causal curves cross each horizon. Together with the strict monotonicity of $r$ in the intervening $\Delta<0$ regions, Eq.~\eqref{eq:one} excludes any future-directed causal curve from crossing a horizon back into $\mathscr E$. Thus no causal signal from either interior can return to $\mathscr E$ and subsequently reach $\mathscr I$. Since every regular point of $\mathscr E$ can reach $\mathscr I$, we obtain
\begin{equation}
\mathcal H^+=\partial J^-(\mathscr I)
=\mathcal H_1^+\cup\mathcal H_2^+\,.
\label{eq:eh}
\end{equation}
Hence $\mathcal H_1^+$ and $\mathcal H_2^+$ are precisely the two disconnected components of the future event horizon, defining two black-hole regions $\mathcal B_1$ and $\mathcal B_2$, respectively.

\begin{figure}[htbp]
        \centering
        \includegraphics[width=0.98\linewidth]{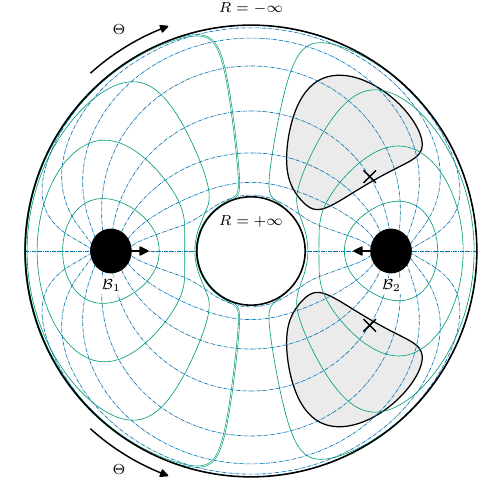}
        \caption{Topological representation of the fixed-$(\tau,\Phi)$ section. For physical intuition, the black-hole rods are represented as black disks, while the remainder of the diagram is geometrically faithful. Curve styles, symbols, and shading are as in Fig.~\ref{fig:weyl-orbit}. The arrows attached to the black holes indicate their rotation axes. The diagram consists of two identical copies of the Weyl diagram, glued along the regular segments of the rotation axis to make the global topology manifest.}
        \label{fig:annulus}
\end{figure}

The common exterior $\mathscr{E}$ of the two black holes has nontrivial topology. To see this geometrically, return to the old coordinates and temporarily ignore the two $\Omega=0$ points and the ring singularity. On a constant-$t$ slice, consider a constant-$r$ two-sphere enclosing black hole $\mathcal{B}_1$. As $r$ increases, the sphere sweeps through the exterior until it reaches the regular continuation surface $r=+\infty\sim-\infty$. Passing into the neighboring patch and continuing with $r$ increasing from $-\infty$, the same family of spheres eventually encloses black hole $\mathcal{B}_2$. Such a continuous sweep is impossible in ordinary $\mathbb{R}^3$ with two three-balls removed, and instead reflects the topology of $S^2\times[0,1]$, or equivalently $S^3$ with two three-balls removed. Restoring now the two $\Omega=0$ loci, which are blown up into complete BR null infinities in the new coordinates, together with the ring singularity, whose removal amounts to excising a tubular neighborhood, a constant-$t$ section of $\mathscr{E}$ is therefore topologically $S^3$ with four three-balls and one solid torus removed. Accordingly, its homology groups are $H_1(\mathscr{E};\mathbb{R})\simeq\mathbb{R}$ and $H_2(\mathscr{E};\mathbb{R})\simeq\mathbb{R}^4$. Fig.~\ref{fig:annulus} shows a complementary illustration.

\begin{figure}[htbp]
\centering
\includegraphics[width=0.98\linewidth]{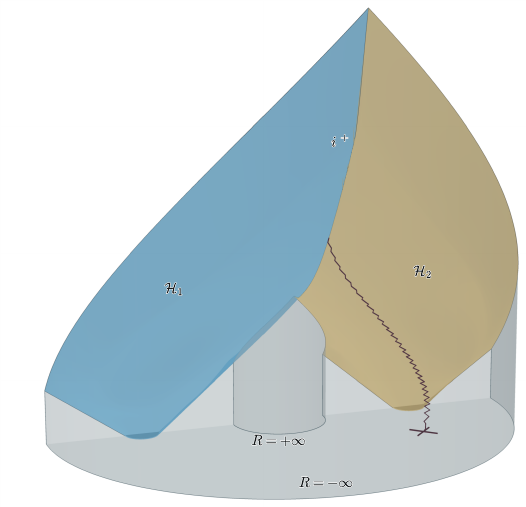}
\caption{Three-dimensional schematic illustration of the causal structure of the extended KNBR spacetime, showing only its future half. The base represents one half of the annulus in Fig.~\ref{fig:annulus}. The two translucent cylindrical boundaries represent the null infinities at $R\to\pm\infty$. The blue and yellow surfaces represent the future horizon components $\mathcal H_1^+$ and $\mathcal H_2^+$, respectively. The two black holes share a connected exterior. The jagged curve represents the ring curvature singularity. The upper intersection of the two horizon surfaces, together with the upper intersection curves of the horizons with the $R\to\pm\infty$ boundaries, all represent the future timelike infinity $i^+$ of the common exterior.
}
    \label{fig:pen}
\end{figure}

The causal structure of the extended spacetime is illustrated by the three-dimensional schematic diagram in Fig.~\ref{fig:pen}. Following the schematic approach of~\cite{Griffiths:2006tk}, we use the orbit-space representation as the base to display the horizons, the two null infinities, and the ring singularity within a single diagram.

\section{Conclusion and Discussion}
We have introduced a new coordinate system for the extended KNBR spacetime that incorporates the previous reciprocal continuation across $r=\infty$ into a single patch description. The two loci at $\Omega=0$ are resolved into complete BR null infinities. The resulting Weyl description makes the global structure of the spacetime manifest. In the regular static limit, the same coordinates also identify the spacetime globally with the balanced Alekseev--Garc\'ia geometry as shown in Appendix~\ref{sec:static}. 
%\z{\sout{ and clarify how the reciprocal continuation across Schwarzschild-like radial infinity is embedded in its global structure}}.

Remarkably, the two future Killing horizons form two disconnected components of the future event horizon, describing two black holes that share a common exterior. The ring singularity is causally visible from the null infinities and is therefore naked. The common exterior further possesses nontrivial spatial topology. The extended KNBR geometry thus provides an unusual exact stationary spacetime in which two black holes and a naked singularity coexist within a topologically nontrivial exterior with BR null infinities.

Our discussion has focused primarily on the common exterior region between the two black holes. The geometry inside each horizon is expected to retain the local structure found in the previous extensions of $\text{KBR}_\text{s}$ and KNBR~\cite{Zhou:2026tkm,Ovcharenko:2026ooh,Zhou:2026fvj}. However, the previous global landscape, in which the interiors of the two black holes were glued to each other, should be abandoned. For the present two-black-hole interpretation, these interiors remain distinct and may admit independent global completions. A systematic analysis of these possible interior completions is left for future work.

The presence of a naked ring singularity also raises a broader question concerning the weak cosmic censorship conjecture. Standard mathematical formulations of the conjecture are usually posed for generic regular asymptotically flat initial data and formulated in terms of the completeness of future null infinity. The KNBR spacetime lies outside this setting, since its asymptotic regions are of BR type. The present solution suggests that a more general formulation may need to accommodate different asymptotic structures and define censorship relative to the corresponding asymptotic observers. Whether such a naked singularity is physically relevant is therefore closely tied to the stability of the spacetime.

The present spacetime is also likely to be unstable. Although $\text{KBR}_\text{s}$ is a special member of the KNBR family, its relevant physical properties are not qualitatively different from those of generic KNBR spacetimes. Previous analyses of the $\text{KBR}_\text{s}$ quasinormal modes~\cite{Zhou:2026tkm,Zhou:2026fvj} established the instability under test scalar perturbations. There, requiring $\Psi/\Omega$ to be regular at the $\Omega=0$ loci is equivalent to imposing Dirichlet boundary conditions at the BR null infinities identified here. The quasinormal-mode problem is therefore unchanged by blowing up the $\Omega=0$ loci, and the spectrum remains the same. The instability has been attributed to the chronology-violating region and to a black-hole-bomb mechanism, both of which are expected to persist in generic KNBR spacetimes.

A further indication comes from the equilibrium of the stationary configuration. This raises the natural question of how the two black holes remain balanced against their mutual gravitational attraction. The balance does not appear to be sustained by electromagnetic repulsion and is not supported by conical defects, but instead seems to arise from the highly symmetric global configuration shown in Fig.~\ref{fig:annulus}. It may therefore resemble a ball balanced at the top of a hill, perfectly stationary in the symmetric configuration but potentially unstable against perturbations that displace it. Moreover, the two black holes generally have different Hawking temperatures, suggesting a further possible thermodynamic instability.

Multi-black-hole solutions with BR asymptotics have also been constructed recently~\cite{Furugori:2026pdt,Clement:2026vzr}. These constructions are restricted to extremal multi-center configurations and include both static Majumdar--Papapetrou-type solutions~\cite{Majumdar:1947eu,Papapetrou:1947} and more general stationary Israel--Wilson--Perj\'es-type solutions~\cite{Perjes:1971gv,Israel:1972vx}. The static subclass admits a natural near-horizon/decoupling interpretation, since it can be obtained from clusters of extremal Reissner--Nordstr\"om black holes by removing the asymptotically flat region, with the resulting BR region representing their common throat~\cite{Maldacena:1998uz}. The KNBR geometry is qualitatively different because its black holes need not be extremal. It therefore provides a considerably broader setting for multi-black-hole geometries with BR asymptotics and may hint at a related decoupling interpretation, although whether an asymptotically flat completion exists remains open.

The coordinate construction introduced here may also be useful for a considerably broader class of spacetimes. $\text{KBR}_\text{s}$ belongs to a broader family of type-D Einstein--Maxwell solutions with nonaligned electromagnetic fields~\cite{Ovcharenko:2025cpm}, whose non-twisting sector includes Schwarzschild and C-metric black holes in a BR background, as well as Reissner--Nordstr\"om black holes accelerating in an external BR field~\cite{Ovcharenko:2026byw}. A new Ricci-flat rotating geometry was subsequently obtained by demagnetizing $\text{KBR}_\text{s}$~\cite{Ma:2026otg}. More recently, the Einstein--Maxwell family has been further generalized to solutions with fully nonaligned electromagnetic fields~\cite{Furugori:2026qbb}. The coordinate structures of some of these BR-related geometries suggest that the conventional radial infinity may not always represent the genuine asymptotic boundary. It would therefore be interesting to investigate whether the present coordinates, or suitable generalizations of this construction, can reveal the maximal extensions and asymptotic structures of these spacetimes.

%\z{\begin{center}
%\boxed{\text{\textbf{Draft below}}}
%\end{center}}

\section*{Acknowledgement}
This work is supported by the National Key R\&D Program of China Grant No. 2022YFC2204603, and by the National Natural Science Foundation of China with grants No. 12475063, No. 12247103, No. 12588101, and No. 12535002.

\appendix
\section{The static limit and the Alekseev--Garc\'ia geometry}
\label{sec:static}
In this appendix, we consider the nonrotating $a\to0$ limit of KNBR, show that its regular limit reduces to the Schwarzschild--BR (SBR) geometry, and demonstrate that the new coordinates provide a global diffeomorphism to the Alekseev--Garc\'ia (AG) solution~\cite{Alekseev:1996fq}. For fixed nonzero $e$ and $B$, this limit is genuinely singular at the level of the geometry~\cite{Ovcharenko:2026tos}. In particular, the BR radius scales as $L=\mathcal{O}(a^2)$, while at either BR end
\begin{equation}
R_{ab}R^{ab}\xrightarrow{R\to\pm\infty}\frac{4}{L^4}=\mathcal{O}(a^{-8})\,.
\label{eq:sl}
\end{equation}
A regular static limit is nevertheless obtained by setting $e=\gamma a$ with finite $\gamma$ and taking $a\to0$. Under the radial redefinition
\begin{equation}
\frac{1}{\bar r}=\frac{1}{r}+B\gamma\,,
\label{eq:rl}
\end{equation}
the metric reduces smoothly to the SBR geometry. The new coordinates also admit a regular limit, which, up to the reflection $R\to-R$, is
\begin{align}
\tau&=BD_0t,&R&=\frac{B(D_0\bar r-m)\cos\theta}{\bar\Omega}\,,\nonumber\\
\cos\Theta&=\frac{1}{\bar\Omega},&\Phi&=\phi\,,
\label{eq:sbr}
\end{align}
where $D_0=1+m^2B^2$, $\bar\Delta=D_0\bar r^2-2m\bar r$, and $\bar\Omega^2=1+B^2\bar\Delta\sin^2\theta$. Here $\bar\Omega$ denotes the smooth signed branch, positive in the original exterior and negative after continuation across $\bar r=\infty$.

The metric can be written entirely in the new coordinates. It is useful to introduce
\begin{align}
\mathcal R_\pm
&=\frac{1}{L^2}\sqrt{\varrho^2+\left(z\mp L^2Bm\right)^2}\geq0\,,\nonumber\\
\mathcal S&=\mathcal R_++\mathcal R_-\,,
\nonumber\\
\mathcal U&=\frac{(\mathcal S-2Bm\cos\Theta)^2}{\mathcal S^2-4B^2m^2}\,,
\nonumber\\
\mathcal V&=\frac{(\mathcal S+2Bm\cos\Theta)^2}{4\mathcal R_+\mathcal R_-}\,.
\label{eq:uv}
\end{align}
The metric then takes the form
\begin{align}
\mathrm{d}s^2&=\frac{1}{B^2D_0^2}\Bigg[-\mathcal U(1+R^2)\mathrm{d}\tau^2+\mathcal V\left(\frac{\mathrm{d}R^2}{1+R^2}+\mathrm{d}\Theta^2\right)\nonumber\\
&+\mathcal U^{-1}\sin^2\Theta\mathrm{d}\Phi^2\Bigg]\,.
\label{eq:sm}
\end{align}
The inverse transformation is
\begin{equation}
\bar r=\frac{\mathcal S+2Bm\cos\Theta}{2BD_0\cos\Theta},\quad \cos\theta=\frac{2R}{\mathcal S}\,.
\label{eq:sinv}
\end{equation}
In particular, near $|\bar r|=\infty$ one has $\cos\Theta\propto\bar r^{-1}$. Thus $\Theta=\pi/2$ is an interior joining surface, and crossing it is precisely the reciprocal continuation obtained by introducing $y\propto1/\bar r$ and extending through $y=0$.

This static geometry is directly related to the solution constructed by Alekseev and Garc\'ia~\cite{Alekseev:1996fq}, which describes a Schwarzschild black hole immersed in the homogeneous electromagnetic field of the BR universe. The relation between the balanced AG solution and the SBR solution was discussed in~\cite{Barrientos:2026shy,Ovcharenko:2026byw} and was recently established locally by an explicit diffeomorphism~\cite{Astorino:2026nhd}, while its global identification remained unresolved. We add hats to their coordinates and parameters whenever they may be confused with ours. In their BR-adapted coordinates, the metric is~\cite{Alekseev:1996fq,Ortaggio:2018ikt}
\begin{align}
\mathrm{d}s^2&=-e^{2\hat\psi}\cosh^2\frac{\hat z}{\hat b}\mathrm{d}\hat t^2+e^{2\hat\gamma}\left(\mathrm{d}\hat z^2+\mathrm{d}\hat\rho^2\right)\nonumber\\
&+e^{-2\hat\psi}\hat b^2\sin^2\frac{\hat\rho}{\hat b}\mathrm{d}\hat\phi^2\,,
\label{eq:ag}
\end{align}
where
\begin{align*}
e^{2\hat\psi}&=\frac{\left(\hat R_++\hat R_--2\hat m\cos(\hat\rho/\hat b)\right)^2}{(\hat R_++\hat R_-)^2-4\hat m^2},\nonumber\\
e^{2\hat\gamma}&=\frac{\left(\hat R_++\hat R_--2\hat m\cos(\hat\rho/\hat b)\right)^2}{4\hat R_+\hat R_-}\nonumber\\
&\times\left[\frac{\hat R_+-\hat b\sinh(\hat z/\hat b)+(\hat l+\hat m)\cos(\hat\rho/\hat b)}{\hat R_--\hat b\sinh(\hat z/\hat b)+(\hat l-\hat m)\cos(\hat\rho/\hat b)}\right]^2\,,
\end{align*}
with
\begin{equation*}
\hat R_\pm^2=\left(\hat l\pm\hat m-\hat b\sinh\frac{\hat z}{\hat b}\cos\frac{\hat\rho}{\hat b}\right)^2+\hat b^2\cosh^2\frac{\hat z}{\hat b}\sin^2\frac{\hat\rho}{\hat b}\,.
\end{equation*}
Here $-\infty<\hat z<\infty$ and $0\leq\hat\rho\leq\pi\hat b$, so that the pair $(\hat\rho,\hat\phi)$ parametrizes a complete BR $S^2$, while $\hat z$ runs along the noncompact $\mathrm{AdS}_2$ direction.

The parameter $\hat l$ determines the displacement of the black hole from the equilibrium position of the BR static frame. For $\hat l\neq0$, a conical defect appears~\cite{Ortaggio:2018ikt}, and $\hat l$ has also been associated with acceleration due to this defect~\cite{Ovcharenko:2026byw}. By contrast, the KNBR geometry has no conical defects, while the limiting metric is invariant under $R\to-R$, which exchanges the two null infinities. Moreover, the KNBR horizon rods are centered at $R=0$, so no displacement along the BR direction is expected to emerge in the static limit. These properties naturally suggest the balanced AG branch $\hat l=0$, as confirmed by the explicit identification below
\begin{align} 
\hat t&=\frac{\tau}{BD_0}\,, &\hat\phi&=\Phi\,, &R&=\sinh\frac{\hat z}{\hat b}\,,\quad \Theta=\frac{\hat\rho}{\hat b}\,,\nonumber\\ \hat b&=\frac{1}{BD_0}\,, &\hat m&=\frac{m}{D_0}\,, &\hat l&=0\,. \label{eq:map} 
\end{align}
Under this map, one has
\begin{equation}
\hat R_\pm=\hat b\mathcal R_\pm\,,\quad
e^{2\hat\psi}=\mathcal U\,,\quad e^{2\hat\gamma}=\mathcal V\,,
\label{eq:id}
\end{equation}
so that \eqref{eq:ag} reduces exactly to \eqref{eq:sm}. A direct calculation shows that the Maxwell field also agrees exactly with the AG field. 

The joining surface is parametrized by
\begin{equation}
\Theta=\frac{\pi}{2}\,,\qquad R=\sqrt{D_0}\cot\theta\,.
\label{eq:se}
\end{equation}
In the AG coordinates, it corresponds to
\begin{equation}
\hat\rho=\frac{\pi\hat b}{2}\,,
\label{eq:eq}
\end{equation}
the equator of the BR $S^2$. Thus the original exterior occupies the northern hemisphere, while continuation across $\bar r=\infty$ passes through the equator into the southern hemisphere toward the antipodal pole.

The SBR horizon is
\begin{equation}
\Theta=0\,,\quad |R|\leq Bm\,,
\label{eq:hp}
\end{equation}
whereas on the continued sheet $\bar r\to0$ corresponds to
\begin{equation}
\Theta=\pi\,,\quad |R|\leq Bm\,.
\label{eq:sp}
\end{equation}
The latter is the antipodal timelike curvature singularity of the balanced AG solution~\cite{Alekseev:1996fq,Ortaggio:2018ikt}, corresponding to $\bar r=0$ of the neighboring Schwarzschild-like patch. Its negative-mass interpretation follows naturally from the discrete reflection
\begin{equation}
(m,e,r)\longrightarrow(-m,-e,-r)\,,
\label{eq:mr}
\end{equation}
of the underlying Einstein--Maxwell family. In the static geometry, this maps the continued negative-$\bar r$ region to a description with positive radial coordinate and opposite mass parameter, reproducing the negative-mass interpretation of the AG singularity.

Meanwhile,
\begin{equation}
R\to\pm\infty\quad\Longleftrightarrow\quad \hat z\to\pm\infty\,,
\label{eq:si}
\end{equation}
so the two BR null infinities resolved by our coordinates are exactly the two asymptotic ends of the balanced AG cylinder.

The rotating family provides a complementary interpretation of the antipodal locus. Defining $k_0=\gamma-mB$, the horizon radius of $\mathcal B_2$ behaves as
\begin{equation}
r_2=\frac{1+k_0^2}{2m}a^2+\mathcal O(a^4),
\label{eq:hr}
\end{equation}
so that $\bar r_2\to0$, while the horizon of $\mathcal B_1$ approaches $\bar r_1=2m/D_0$. Their images therefore approach the loci \eqref{eq:sp} and \eqref{eq:hp}, respectively. In the static limit, $\mathcal B_1$ becomes the Schwarzschild--BR black hole, whereas $\mathcal B_2$ degenerates into the antipodal curvature singularity.

This degeneration of $\mathcal B_2$ is also reflected in its local horizon geometry. Evaluating the horizon area and the surface gravity gives
\begin{equation}
\mathcal A_2=\frac{4\pi}{D_0}a^2+\mathcal O(a^4)\,,\quad \kappa_2=\frac{m}{a^2}+\mathcal O(1)\,.
\label{eq:du}
\end{equation}
Hence the area of $\mathcal B_2$ shrinks to zero while the magnitude of its surface gravity diverges.

Altogether, the new coordinates identify the regular static limit globally with the balanced AG geometry and make explicit how the reciprocal continuation across Schwarzschild-like radial infinity is embedded in its global structure.

\section{Special cases and continuous limits}
\label{sec:special}
We now examine the range of applicability of the new coordinate transformation~\eqref{eq:tr} across the remaining parameter regimes. The metric is invariant under
\begin{align*}
(m,e,r)&\longrightarrow(-m,-e,-r)\,,\\
(a,e,\phi)&\longrightarrow(-a,-e,-\phi)\,,\\
(B,e)&\longrightarrow(-B,-e)\,,\\
(e,s)&\longrightarrow(-e,-s)\,.
\end{align*}
Restricting to the real branch $s^2\geq0$, these discrete symmetries allow us to choose
\begin{equation}
m\geq0\,,\quad a\geq0\,,\quad B\geq0\,,\quad e\in\mathbb{R}\,,\quad s\geq0\,.
\end{equation}

The special case $s=0$ selects a locus in the $(m,a,B,e)$ parameter space, as follows from~\eqref{eq:parameters}. For $e\neq0$, setting $s=0$ does not by itself make~\eqref{eq:tr} singular, while for $e=0$ one has $P_0=0$ and the compact angular sector degenerates, which is unphysical. We therefore restrict to $s>0$. 

For $B=0$, the geometry reduces to Kerr--Newman. The present coordinates are therefore no longer adapted to the geometry as the BR length scale diverges and the BR asymptotic region disappears. Accordingly, the following analysis assumes $B>0$. 

The nonrotating $a\to0$ limit is discussed in Appendix~\ref{sec:static}. Hereafter, we focus on the rotating case $a>0$.

At $e=0$, the apparent singularity of~\eqref{eq:tr} is removable as the transformation admits the continuous limit
\begin{align}
\tau&=BD_0t,&R&=B\sqrt{P_0D_0}\left(r-\frac{m}{D_0}\right)\frac{\cos\theta}{\Omega}\nonumber\\
\cos\Theta&=\frac{s}{\Omega},&\Phi&=\phi+aB^2D_0^2t\,.
\end{align}
Thus the same coordinates apply to the genuine uncharged $\text{KBR}_0$ geometry~\cite{Ovcharenko:2026tos}. In this limit the two $\Omega=0$ points move to axial $r=\infty$, where the reciprocal coordinate $y\propto1/r$ fails to provide a regular chart, yet the new coordinates remain regular. If $m=0$ as well, the spacetime reduces to the BR background, and the new coordinates reduce, after constant rescalings, to the BR coordinates introduced in~\cite{Podolsky:2025tle}. The discussion below concerns the case $e\neq0$. 

We now consider the remaining parameter ranges together, where three additional features arise.

First, for $-B^{-2}<\mu^2<0$, where the lower bound follows from the requirement $P_0>0$, one necessarily has $D>0$. In this range, the two loci $R=0$, corresponding to $r=r_{\rm c}$ and $\theta=\pi/2$, intersect at $(r,\theta)=(r_{\rm c},\pi/2)$ whenever $\Delta'(r_\Omega)\neq0$. For $r_{\rm c}\neq0$, the two branches intersect at a regular point where the coordinate map degenerates, and the original $(r,\theta)$ coordinates provide a complementary patch. If $r_{\rm c}=0$, their only intersection is the ring singularity, so no additional patch is required on the regular manifold.

Second, the behavior at $r=\infty$ changes for $D\leq0$. For $D<0$, the real $(R,\Theta)$ coordinates do not provide a chart of $r=\infty$. For $D=0$ with $m>0$, the $r=-\infty$ side is reached as a boundary of the real coordinate domain. In the special case $m=D=0$, one has $\Delta'(r_\Omega)=0$, but the $(R,\Theta)$ coordinates remain well behaved in the limiting form discussed later. The entire $r=\infty$ hypersurface is then compressed to a single point in the $(R,\Theta)$ orbit space, where the Jacobian of the transformation vanishes.

Third, when $\Delta'(r_\Omega)=0$, the apparent divergence of $r_{\rm c}$ and $E$ is removable. Their combination entering $R$ remains finite and, up to $R\to-R$, becomes
$$
R=\frac{B\sqrt{P_0\Delta(r_\Omega)}\cos\theta}{\Omega}.
$$
This cannot occur in the generic $D>0$, $\mu^2>0$ sector, since
$$
[\Delta'(r_\Omega)]^2=4\left[D\Delta(r_\Omega)+\mu^2\right]>0
$$
and $\Delta(r_\Omega)>0$.

The case $e<0$ does not affect the validity of the coordinate construction but may interchange the assignments of the horizons to $\Theta=0$ and $\Theta=\pi$. With $\xi>0$,
$$
\cos\Theta\big|_{r_1}=-\cos\Theta\big|_{r_2}=-\operatorname{sgn}\left[e\Delta'(r_\Omega)\right]\,,
$$ 
thus the assignments are reversed when $e\Delta'(r_\Omega)>0$.

In the region $\Delta(r)<0$, the surface $r=\infty$ can likewise be covered by analytically continuing the angular coordinate as
\begin{equation}
\Theta=\mathrm{i}\eta\quad\text{or}\quad\Theta=\pi-\mathrm{i}\eta\,,\quad \eta\geq0\,.
\label{eq:hc}
\end{equation}
In particular, Eq.~\eqref{eq:sth} becomes
\begin{equation}
\sinh^2\eta=-\frac{B^2\Delta\sin^2\theta}{\Omega^2}\,.
\label{eq:sh}
\end{equation}
The inverse transformation and the metric in this region follow directly from Eqs.~\eqref{eq:inv} and~\eqref{eq:mRT} under the replacements~\eqref{eq:hc}.
Likewise, the equation determining $u=\xi^2$ follows from its $\Delta>0$ counterpart under the same continuation, with the appropriate root selected by continuity along each branch rather than by the larger-root prescription. Since $u$ determines only $\xi^2$, the sign of $\xi$ is fixed separately by continuity along each branch. Its discriminant is
\begin{equation}
\mathcal D_{\rm u}=\frac{B^4}{\Omega^4}\left(P\Delta+\mu^2I\sin^2\theta\right)^2\,.
\label{eq:hd}
\end{equation}
The orbit-space Jacobian is
\begin{equation}
\left(\det\frac{\partial(R,\eta)}{\partial(r,\theta)}\right)^2=\frac{(1+R^2)\mathcal D_{\rm u}}{-IP\Delta}\,.
\label{eq:hjs}
\end{equation}
Thus, within the $\Delta<0$ region, the coordinate transformation degenerates precisely at $\mathcal D_{\rm u}=0$, where the two roots for $u$ coincide and the map generically develops a fold. Away from the fold, the two roots are distinct, so the same $(R,\eta)$ coordinates correspond to two distinct spacetime points. By unfolding the fold and treating the two roots as separate coordinate branches, the construction can be continued locally across this locus and extended to a larger spacetime region~\cite{Whitney:1955,Golubitsky:1973}.

Taken together, these cases show that the new coordinate construction remains applicable across a broad range of the parameter space.

\bibliography{mainRef}
\bibliographystyle{apsrev4-1}

\end{document}